\documentclass[twocolumns]{aa} 

\usepackage[dvipsnames]{xcolor}
\usepackage{amssymb}
\usepackage{natbib}
\usepackage{graphicx}
\usepackage{txfonts}
\usepackage[normalem]{ulem}
\usepackage{float}
\usepackage[caption = false]{subfig}
\usepackage{array}

\begin{document}

\sloppy
  \title{Evolution of long-lived globular cluster stars}
   \subtitle{IV. Initial helium content and white-dwarf properties} 

\authorrunning{W. Chantereau et al.} \titlerunning{IV. Initial helium content and white dwarf properties}   

   \author{W. Chantereau \inst{1}
     \fnmsep\thanks{E-mail: william.chantereau@unige.ch}
          \and C. Charbonnel \inst{1,2} 
          \and G. Meynet \inst{1} 
        }

   \institute{Department of Astronomy, University of Geneva, Chemin des Maillettes 51, CH-1290 Versoix, Switzerland
         \and
         IRAP, UMR 5277 CNRS and Universit\'e de Toulouse, 14 Av. E. Belin, F-31400 Toulouse, France}
   
  \date{}
 
  \abstract
  {Globular clusters host stars with chemical peculiarities. The associated helium enrichment is expected to affect the evolution of stars, in general, and of low-mass stars, and in particular the progenitors of white dwarfs (WDs).}
  {We investigate the effects of different initial helium contents on the properties of white dwarfs such as their masses, compositions, and the time since their formation.}
  {We used the grid of stellar models that we presented in the first papers of this series, which were computed for low-mass, low-metallicity stars with different helium content at [Fe/H]=-1.75 up to the end of the thermally pulsing asymptotic giant branch (TP-AGB) phase. We determined an initial-to-final mass relation as a function of the initial helium mass fraction, where the final mass is determined at the end of the TP-AGB phase. We couple the results with different possible distributions of the initial helium content for low-mass stars in NGC~6752 to predict the properties of WDs in this cluster.}
  {In a globular cluster at a given age, the He enrichment implies lower initial masses for stars at a given phase. Thus it leads to a decrease of the masses of white dwarfs reaching the cooling sequence. In addition the He enrichment increases the total mass and number of white dwarfs and eventually allows the presence of He white dwarf from single progenitors.}
  {The low He enrichment determined in most globular clusters with different methods results in negligible effects on the white dwarf properties. However, in the few globular clusters that display a high He enrichment, this may significantly affect the characteristics of the white dwarfs. In NGC~2808 and $\omega$ Centauri the high He enrichment even leads to the formation of He white dwarfs from single He-rich progenitors. Therefore investigating the white dwarf mass domain in globular clusters with a high He enrichment would provide an additional indirect way to measure and constrain the He enrichment degree.}


        \maketitle
\section{Introduction}\label{intro}

In the last few decades spectroscopic and photometric observations have shown that globular clusters (GCs) are not consistent with simple stellar populations \citep[e.g.,][]{Carretta09,Piotto09,Gratton12}. These observations are usually related to the presence of multiple populations, in which long disappeared massive stars of a first population (1P) of stars polluted the intra-cluster medium. In this environment a second population (2P) of low-mass stars formed with a chemical composition that is typical of the ejecta of the 1P polluting stars (He-, N-, Na-, Al-enriched, and C-, O-, and Mg-depleted) diluted with intra-cluster matter. The different scenarios that investigate the role of various types of possible 1P polluting stars are still strongly debated \citep[see, e.g.,][]{Bastian15,Renzini15,Charbonnel16_EES}. The most invoked scenarios in the literature are the asymptotic giant branch scenario \citep[AGB;][]{Ventura01} and the fast rotating massive stars scenario \citep[FRMS;][]{Decressin07}. These scenarios differ especially in the prediction of the amplitude of helium enrichment for 2P stars. In the AGB and FRMS scenarios, a minimal helium content as low as $\sim$0.3 allows to reproduce the bulk of the abundance observations but not the extreme abundances observed among 2P stars. The AGB scenario leads to a maximal helium content of $\sim$0.36-0.38 in mass fraction \citep{Ventura13,Doherty14}, whereas the FRMS scenario in its current form provides a maximum value of 0.8.

The aim of this series of papers \citep[][paper I, II, and III respectively]{Chantereau15,Charbonnel16,Chantereau16} is to explore the effects of the helium enrichment in GCs within the framework of the FRMS scenario. Since a high helium content has an important effect on the evolution of 2P low-mass stars \citep[e.g.,][]{Salaris06,Pietrinferni09,Sbordone11,Valcarce12}, we quantified its effects on different properties of GC stars at different ages, from the zero-age main sequence up to the end of the AGB. 

White dwarfs (WDs) are the remnants of low- and intermediate-mass stars. As long as they remain in their parent GC they provide a fantastic window on the integrated population of their progenitors over the whole GC lifetime. White dwarfs provide an additional way to determine several properties of the cluster they are populating. For instance WDs allow us to infer the distance of the cluster and its age \citep[see, e.g., the review from][]{Moehler08}. Therefore WDs can be very useful tools provided that their masses and spectral type are well known. Thus the next logical step in this series of papers is to consider helium effects on the properties of WDs. 

Many studies have been devoted to GC WDs \citep[e.g.,][]{Renzini96,Zoccali01,Hansen03_review,Hansen03,Moehler04,Bedin05,Strickler09,Richer13,GarciaBerro14,Torres15}. So far, however, only \cite{Althaus16} have investigated in detail some effects of He enrichment on WD properties \citep[see also, to a certain extent,][]{Cassisi09}. 

Deriving the He content of stars is a challenging task. In the case of GC stars, direct measurements are rare \citep{Moehler07,Villanova09,Pasquini11,Villanova12,Dupree13,Gratton14,Marino14,Mucciarelli14,Gratton15}. Several studies use instead indirect methods based, for example, on isochrone fitting using high precision photometry \citep[see, e.g.,][]{Piotto07,Anderson09,Bragaglia10_2808,diCriscienzo10_6397,diCriscienzo10,King12,Lee13,Milone15_7089}. As of today, for most of the GCs under scrutiny only modest He enrichments were derived with respect to the typical Galactic chemical enrichment (a helium mass fraction increase of $\sim$0.02-0.03). However there are a few GCs that display high He enrichment (a helium mass fraction increase higher than $\sim$0.1), such as NGC~2419, NGC~2808, $\omega$ Centauri, and NGC~6388 \citep{Busso07,King12,diCriscienzo15,Milone15}. Thus in these GCs the evolution of 2P stars and in turn their WD properties are greatly affected by the helium content. For instance observed He-WDs are attributed to a close binary evolution; this is the case because otherwise only very low-mass single stars (M$_\mathrm{ini} \lesssim$ 0.7~M$_\odot$) with lifetimes longer than the age of the Universe can provide such WDs. However single He-rich stars have significantly shorter lifetimes and thus may produce observed He-WDs in GCs, such as  NGC~6791 and $\omega$ Centauri \citep[e.g.,][]{Hansen05,Calamida08,Cassisi09,Bellini13}.

In this last paper of the series, we investigate in a fully self-consistent way the effects of helium on the properties (mass, age, and composition) of the WDs that originate from 2P progenitors. It will allow us to infer whether He enrichment in GCs plays a significant role in WD characteristics that may affect, for instance, the dynamical properties of GCs. After a brief presentation of our hypothesis and the definitions used in this paper (Sect.~\ref{input}), we investigate how varying the initial He content for a star with a given initial mass changes the mass of the resulting WD and compare our results with other studies (Sect.~\ref{08msun}). Then, we explore the effects of initial mass at an initial given helium content and determine a corresponding initial-to-final mass relation at a low metallicity in the framework of the multiple populations phenomenon (Sect.~\ref{main}). In Sect.~\ref{wdproperties} we investigate the WD properties (mass range and composition) as a function of the age of their host GC if we assume an initial helium distribution predicted by the FRMS scenario. We then compare our predictions with the masses of the WDs and He enrichment determined in GCs (Sect.~\ref{compobs}). Finally, we summarize our results in Sect.~\ref{Conclusions}.  

\section{Hypothesis and definitions}\label{input}
\subsection{Input physics of the stellar models}

We use the grid composed of 420 stellar evolution models presented in Papers I and II. These standard (no rotation, atomic diffusion, and overshooting) models were computed from the pre-main sequence up to the post-AGB phase with the stellar evolution code STAREVOL \citep[e.g.,][]{Forestini97,Siess00,Lagarde12}. We extensively exploit the predictions of the grid of models computed (initial stellar masses range between 0.3~M$_\odot$ and 1~M$_\odot$) for [Fe/H] = -1.75 with an alpha-enhancement of 0.3~dex, which is best suited for the case of the well-documented GC NGC~6752 \citep{Carretta10}. We also use models computed with [Fe/H] = -2-2, -1.15, and -0.5 ([$\alpha/Fe$] = +0.35, +0.3, and +0.2 respectively) to discuss the impact of metallicity on our conclusions. 

For the 1P stars, the adopted canonical initial helium mass fraction is $Y_\mathrm{ini}$ = 0.248 at [Fe/H] = -1.75 ($Y_\mathrm{ini}$ = 0.248, 0.249, and 0.255, respectively for the models with [Fe/H] = -2.2, -1.15, and -0.5). In the framework of the FRMS, the initial chemical composition of the 2P low-mass stars (assumed to be only stars with masses lower than 1~M$_\odot$) results from the dilution of H-burning products ejected by the 1P fast rotating massive stars with the intra-cluster matter \citep{Decressin07}. The most extreme 2P stars are predicted to be born with an initial helium mass fraction of 0.8 and the other element abundances change accordingly. 

The models were computed with the Reimers prescription \citep{Reimers75} for mass loss along the evolution up to the end of central He burning, assuming a value of 0.5 for the $\eta$ parameter regardless of the initial He content. This is consistent with the value of $\eta$ derived by \cite{McDonald15}, fitting the median mass of the horizontal branch stars of 56 Galactic GCs (0.477$\pm$0.070). Their value is based on the full cluster population, regardless of the helium content of the different subpopulations, whose impact on the mass-loss rate in RGB stars is not currently understood. During the thermally pulsing asymptotic giant branch phase (TP-AGB), the mass loss is treated in our models with the empirical law relating the mass-loss rate and the period of pulsations occurring during this phase \citep{Vassiliadis93}. 

In line with this series of papers, we focus here on the effects of large variations in the initial He content on the composition (He versus CO) and mass of the resulting white dwarfs. 
It is thus beyond the scope of this study to look at the impact of core overshooting, atomic diffusion, or rotation. Although these effects might slightly modify the resulting WD masses (in a way that might eventually vary with the initial He content), this should not affect our conclusions on the differential importance of He.

\subsection{Definitions}\label{def}

In this study we want to determine the WD mass as it can be deduced from our stellar evolution models as a function of the initial mass (M$_\mathrm{ini}$) and initial helium content ($Y_\mathrm{ini}$) at a given metallicity. Some authors approximate the WD mass by the core mass at the first TP \citep[see, e.g.,][]{Weidemann00}. However the mass growth of the CO core during the TP-AGB phase is not negligible (0.07~M$_\odot$ during the TP-AGB phase for the 0.8~M$_\odot$ computed at [Fe/H] = -1.75 with $Y_\mathrm{ini}$ = 0.248). Moreover, as discussed later (section~\ref{helium} and Table~\ref{table:08Msun}), this mass increase is greater for higher initial helium content. Therefore, our models are computed self-consistently up to the end of the TP-AGB, and the theoretical mass of the WD (M$_\mathrm{WD}$) corresponds to the core mass (M$_\mathrm{core}$), which is the\ mass depth at which the energy production from the helium-burning shell is maximum, on the post-AGB phase at the maximum T$_\mathrm{eff}$. If the star ends its life as He-WD or is a hot-flasher (HF)\footnote{Stars with a delayed helium flash occurring at high effective temperature \citep[see, e.g.,][]{Castellani93,DCruz96,Brown01,MillerBertolami08}.}, we take as WD mass the total stellar mass at the maximum T$_\mathrm{eff}$ after the star crossed the Hertzsprung-Russell diagram (HRD) towards the WD cooling curve.

The ages of the WD (i.e., time since arrival on the WD cooling sequence) given in this study are counted taking the end of the core He-burning phase as a starting point. In a cluster at a given age (Sect.~\ref{wdproperties} and \ref{compobs}), the WD originating from progenitors with initial masses between 0.3 and 1~M$_\odot$ (i.e., in the mass range studied in the present work) are then classified into three categories:

\begin{itemize}
\item Young WDs, that is to say, those that have joined the cooling sequence for less than 500~Myr.
\item Slightly older WDs that have joined the cooling sequence for less than 2~Gyr.
\item All the WDs.
\end{itemize}

This categorization is made to facilitate the comparisons of our theoretical predictions with observations (Sect.~\ref{compobs}). The observations from the literature presented in this paper focus on young and bright WDs. We also separate WDs that have joined the cooling sequence for less than 2~Gyr to provide a category of WDs that cover the entire upper part of the WD cooling curve.  

\section{Effects of helium and metallicity on a 0.8~M$_\odot$ model}\label{08msun}

\subsection{Helium effects} \label{helium}

\begin{table}
\begin{tabular}{c | c | c | c | c | c | c}
        \hline 
    [Fe/H] & $Y_\mathrm{ini}$ & $\tau_\mathrm{end Heb}$ & \multicolumn{2}{c|}{First TP} & \# TP & {\scriptsize Post-AGB}  \\ 
    $Z$ & & & M$_\mathrm{tot}$ & M$_\mathrm{core}$ & & M$_\mathrm{core}$ \\ \hline 
    -2.2 & 0.248 & 13.84 & 0.60 & 0.48 & 7 & 0.57 \\ 
    0.0002 & 0.400 & 5.27 & 0.68 & 0.52 & 22 & 0.67 \\ \hline   
        -1.75 & 0.248 & 14.29 & 0.57 & 0.47 & 8 & 0.54 \\
        0.0005 & 0.260 & 13.18 & 0.58 & 0.47 & 8 & 0.55 \\ 
    & 0.270 & 12.34 & 0.59 & 0.47 & 8 & 0.57 \\
        & 0.300 & 10.35 & 0.61 & 0.48 & 11 & 0.59 \\    
        & 0.330 & 8.55 & 0.63 & 0.49 & 13 & 0.61 \\ 
    & 0.370 & 6.59 & 0.65 & 0.51 & 16 & 0.63 \\
        & 0.400 & 5.40 & 0.66 & 0.53 & 18 & 0.64 \\        
        & 0.425 & 4.56 & 0.67 & 0.55 & 20 & 0.66 \\    
        & 0.450 & 3.84 & 0.68 & 0.57 & 21 & 0.67 \\ 
    & 0.475 & 3.23 & 0.69 & 0.60 & 21 & 0.68 \\
        & 0.525 & 2.26 & 0.71 & 0.64 & 20 & 0.70 \\   
    & 0.600 & 1.31 & 0.73 & 0.71 & 12 & 0.72 \\
    & 0.800 & 0.26 & - & - & 0 & 0.74 \\ \hline
        -1.15 & 0.249 & 16.44 & - & - & HF & 0.50 \\ 
        0.002 & 0.270 & 14.39 & 0.51 & 0.46 & 2 & 0.51 \\ \hline \hline    
    -0.5 & 0.255 & 23.27 & - & - & He-WD & 0.47 \\ 
    0.008 & & & & & & \\ \hline 
    0.0 & 0.276 & 25.58 & - & - & He-WD & 0.46 \\
    0.017 & & & & & & \\    
\hline 
\end{tabular}
\caption[0.8~M$_\odot$]{Age at the end of the core He-burning phase (Gyr), except for the cases at [Fe/H] = -0.5 and 0.0, where the age at the highest luminosity of the RGB phase is shown, for the 0.8~M$_\odot$ as a function of the initial helium content $Y_\mathrm{ini}$ at different metallicities [Fe/H]. The M$_\mathrm{tot}$ and M$_\mathrm{core}$ (M$_\odot$) at the first TP, number of TPs and M$_\mathrm{core}$ on the post-AGB phase at the maximum T$_\mathrm{eff}$. For each metallicity, the first $Y_\mathrm{ini}$ corresponds to the canonical value of helium; HF stands for hot flasher and when the star does not end its life as a CO-WD (when no indication is given) then He-WD is specified.}  \label{table:08Msun}
\end{table}

We focus on the impact of initial helium for a given initial mass on the nature and mass of the WD remnant; for a more detailed discussion of the impact of helium, we refer to Paper I and II. We illustrate this with quantitative values for a 0.8~M$_\odot$ model at [Fe/H] = -1.75 in Table~\ref{table:08Msun}. 
The lifetime of a He-rich star is much shorter than of its He-normal counterpart. For instance the age at the end of the central He-burning phase for $Y_\mathrm{ini}$ = 0.248, 0.4, and 0.8 models are 14.3, 5.4, and 0.26~Gyr, respectively (Table~\ref{table:08Msun} and Fig.~3 in paper I). 

The number of TPs first increases with the helium content up to $Y_\mathrm{ini}$ = 0.475 and then decreases for higher He content. Below $Y_\mathrm{ini}$ = 0.475, owing to shorter lifetimes, the total mass lost of He-rich stars is lower, and in turn their total stellar mass at the start of the TP phase is higher. This leads to a higher mass above the core at the first TP, and thus He-rich models undergo a larger number of TPs than their He-normal counterparts. Above $Y_\mathrm{ini} \sim 0.475$, the increase of the core mass is the dominating effect. Little mass remains above the core and the number of TPs decreases. At a certain point, there is nearly no envelope left above the core (models with $Y_\mathrm{ini} \geq 0.625$). In this case the model does not undergo any TPs and, after the early AGB phase, it directly crosses the HRD towards the maximum T$_\mathrm{eff}$. The values of $Y_\mathrm{ini}$ presented in this section are slightly different from those given in Paper I (sections 4.4 and 4.5) because we use here the more refined grid presented in Paper II, where we added 90 stellar models with $Y_\mathrm{ini}$ = 0.425,0.475,0.525,0.575,0.625, and 0.675.

As displayed in Table~\ref{table:08Msun}, the WD mass of a 0.8~M$_\odot$ stellar model at [Fe/H] = -1.75 for a He-normal content is 0.54~M$_\odot$. This WD mass increases with increasing helium up to 0.74~M$_\odot$ for $Y_\mathrm{ini} = 0.8$, which is an increase of 37~\% of the WD mass with respect to the He-normal case. \\

\subsection{Metallicity effects}

The WD mass of a star for a given initial mass decreases with increasing metallicity. In our models, at [Fe/H] = [-2.2;-1.75;-1.15;-0.5;0.0], the M$_\mathrm{WD}$ are [0.57;0.54;0.50,0.47,0.46]~M$_\odot$  for a canonical initial helium mass fraction and an initial mass of 0.8~M$_\odot$, respectively.\\

Therefore the initial mass of a WD of a given initial mass and helium content is larger in more metal rich clusters than in metal poor clusters.

\subsection{Comparisons between different stellar models}\label{comptheo}

\begin{table*}
\centering
\begin{tabular}{|c | c | p{5cm} | c | c | c | c |}
        \hline 
    Reference & Code & Mass loss & Ov. & $Z$ & $Y_\mathrm{ini}$ & M$_\mathrm{WD}$ \\ \hline
    This study & STAREVOL & Pre-AGB: \cite{Reimers75}, $\eta_\mathrm{R}$ = 0.5 & NO & 0.0005 & $Y_\mathrm{canonical}$ & 0.54 \\
     &  & AGB: \cite{Vassiliadis93} &  &  & 0.4 & 0.64 \\
     &  &  &  & 0.0002 & $Y_\mathrm{canonical}$ & 0.57 \\ 
     &  &  &  &  & 0.4 & 0.67 \\ 
     &  & Pre-AGB: \cite{Schroder05} &  & 0.0005 & $Y_\mathrm{canonical}$ & 0.51 (HF) \\ 
        &  & AGB: \cite{Schroder05,Groenewegen98,Groenewegen09} & & & 0.4 & 0.55 \\ \hline       
    \cite{Bertelli08} & Padova & Pre-AGB: NO & NO & 0.0004 & 0.4 & $>$0.60 \\
    &  & AGB: \cite{Bowen91} & & & & \\ \hline
    \cite{Althaus16} & LPCODE & Pre-AGB: \cite{Schroder05} & YES & 0.0005 & 0.4 & 0.56 \\ 
        &  & AGB: \cite{Schroder05,Groenewegen98,Groenewegen09} & & & & \\ 
\hline 
\end{tabular}
\caption[Theoretical comparisons]{M$_\mathrm{WD}$ from models computed with different stellar evolution codes and different prescriptions for a 0.8~M$_\odot$. "Ov." indicates whether overshoot was included in the computation of the models; HF stands for hot flasher.} \label{theo}
\end{table*}

In this section we compare the WD mass we obtained for the 0.8~M$_\odot$ with predictions of models from the literature (see Table~\ref{theo} for details). \\

At $Z = 0.0004$, \cite{Bertelli08} predicted a core mass at the start of the TP-AGB phase of 0.60~M$_\odot$ for $Y_\mathrm{ini}$ = 0.4 that is larger than our prediction (0.53~M$_\odot$). The difference can be explained by the fact they did not take mass loss during the pre-AGB phases into account.

\cite{Althaus16} obtained a WD mass significantly lower than ours\footnote{In Sect.~4.5 of paper I we specifically defined the core mass as the sum of the CO core mass and the He-burning shell mass and thus found a WD mass for the 0.8~M$_\odot$ model with $Y_\mathrm{ini}$ = 0.4 of 0.66~M$_\odot$ instead; the core mass values present in the models available online correspond to the CO core mass only.} owing to the lower mass-loss rates we are using (especially during the TP-AGB phase). We also tested their mass-loss prescription and found in this case a very slight difference ($\Delta$M$_\mathrm{WD} = $+0.01~M$_\odot$ in their study), which may come from the inclusion of overshoot in their computations. 
To summarize, very similar results are obtained when equivalent physical ingredients are used (here the treatment of the convective zone boundaries, i.e., overshoot and the mass-loss prescription).

\section{Initial-to-final mass relation}\label{main}

\begin{figure}[ht]
   \centering
   \includegraphics[width=0.45\textwidth]{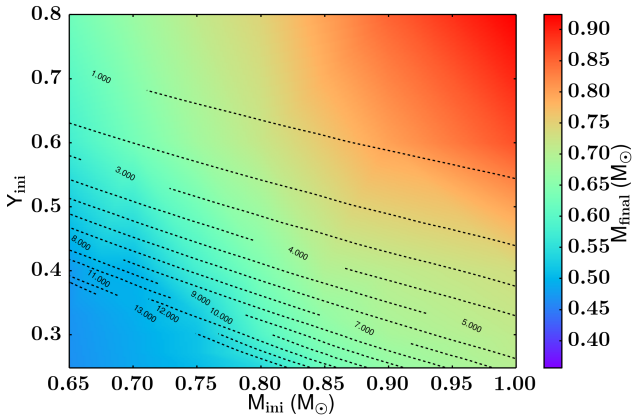}   
    \caption{Final mass of the star (M$_\mathrm{final}$, color-coded) as a function of $Y_\mathrm{ini}$ and M$_\mathrm{ini}$. The dashed lines correspond to isochrones at the end of the central He-burning phase from 1~Gyr to 13~Gyr (shown at each Gyr) at [Fe/H] = -1.75.}
    \label{Figure:Mfinaldiag}
\end{figure} 

\begin{figure}[ht]
   \centering
   \includegraphics[width=0.45\textwidth]{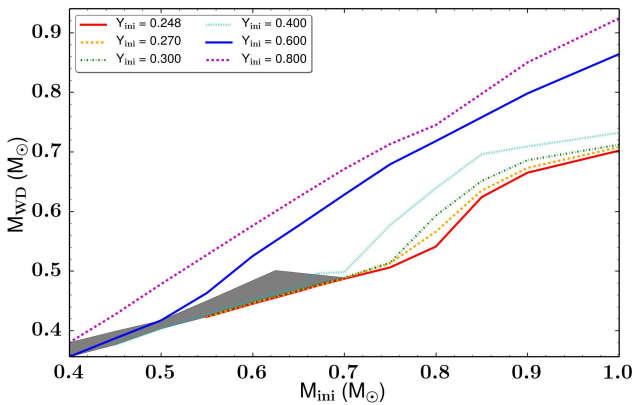}   
    \caption{Mass of the WD as a function of M$_\mathrm{ini}$ for different $Y_\mathrm{ini}$ (colored lines) at [Fe/H] = -1.75. The gray area represents the He-WD domain.}
    \label{Figure:finalmass}
\end{figure} 

We summarize the predictions of our computations for the final mass of the star (i.e., the WD mass) as a function of its initial mass and initial He content for [Fe/H] = -1.75 in Fig.~\ref{Figure:Mfinaldiag}. We show in Fig.~\ref{Figure:finalmass} the initial-to-final mass relations for the different initial helium contents and initial masses at [Fe/H] = -1.75. At canonical $Y_\mathrm{ini}$, the red curve shows an inflection point around 0.8~M$_\odot$. For initial masses above $\sim$0.8~M$_\odot$, stars undergo several TPs leading to a large increase of M$_\mathrm{WD}$. For initial masses below $\sim$0.8~M$_\odot$, the mass of the envelope after the central He-burning phase is very small. Thus there are very few or no TPs and a very slight or no core growth after this phase. This is also the case for lower M$_\mathrm{ini}$, which end their life as He-WDs (gray area in Fig.~\ref{Figure:finalmass}). For higher $Y_\mathrm{ini}$, this inflection point shifts towards lower M$_\mathrm{ini}$ since a higher $Y_\mathrm{ini}$ allows us to ignite central helium for lower mass stars compared to He-normal counterparts. 

To summarize, He-rich models predict that CO-WDs have larger masses and the production of He-WDs from a lower M$_\mathrm{ini}$ than their He-normal counterparts. These differences become important for $Y_\mathrm{ini}$ larger than 0.3.

\section{White dwarf properties at given ages}\label{wdproperties}

In this section we predict the mass domain and composition of WD that we can expect in GCs when taking into account He enrichment (results shown in Table~\ref{table:WD_pred}). We first investigate the WD properties at different ages if we assume that a cluster is made of only He-normal stars  (Sect.~\ref{Predictions}). Then we discuss the expected WD properties of NGC~6752 by adopting the initial helium content distribution predicted by the original FRMS scenario.

\subsection{He-normal stars}\label{Predictions}

\begin{table}
\centering
\begin{tabular}{c | c | c | c | c | c | c}
        \hline 
    $Y_\mathrm{ini}$ & \multicolumn{2}{|c|}{13~Gyr} & \multicolumn{2}{|c}{12~Gyr} & \multicolumn{2}{|c}{9~Gyr} \\ \hline
    & M$_\mathrm{ini}$ & M$_\mathrm{WD}$ & M$_\mathrm{ini}$ & M$_\mathrm{WD}$ & M$_\mathrm{ini}$ & M$_\mathrm{WD}$ \\ \hline
    0.248 & 0.82 & 0.58 & 0.84 & 0.60 & 0.91 & 0.68 \\
    0.270 & 0.79 & 0.56 & 0.81 & 0.59 & 0.88 & 0.65 \\
    0.280 & 0.78 & 0.55 & 0.79 & 0.57 & 0.86 & 0.65 \\
    0.300 & 0.75 & 0.52 & 0.77 & 0.55 & 0.83 & 0.63 \\
    0.400 & 0.63 & 0.47 & 0.64 & 0.47 & 0.70 & 0.50 \\
    0.600 & 0.41 & 0.35 & 0.42 & 0.36 & 0.45 & 0.39 \\
\hline 
\end{tabular}
\caption[WD mass predictions]{M$_\mathrm{ini}$ and M$_\mathrm{WD}$ of stars with an age of 9, 12 and 13~Gyr at the end of the central He-burning phase at [Fe/H] = -1.75.} \label{table:WD_pred}
\end{table}

We first describe the WD characteristics we can expect from our models if we assume only He-normal stars as their progenitors. We focus on clusters with an age of 12 and 13~Gyr (range of ages for NGC~6752).

For the initial stellar mass function (IMF) we use a lognormal mass function \citep{Paresce00} for the stellar masses below 0.85~M$_\odot$ (down to 0.3~M$_\odot$), and a power-law distribution with the \cite{Salpeter55} prescription for more massive stars up to 1~M$_\odot$. We consider an initial population of 300 000 stars to avoid stochastic effects. We do not take into account dynamical effects that could lead to the loss of a fraction of stars by the cluster. Thus the decreasing fraction of stars still alive at a given age is only due to stellar evolution. Finally we do not account for the effects of binaries. \\  

At an age of 13~Gyr, the stars that just reach the WD dwarf stage are those with an initial mass around 0.82~M$_\odot$. Considering the initial mass range investigated in the present work (0.3-1~M$_\odot$), this means that all the stars between 0.82 and 1~M$_\odot$ have evolved into the WD domain. The WDs represent 16~\% of the initial population of 300 000 stars. The WD masses are greater than 0.58~M$_\odot$ and these are only CO-WDs (see Fig.~\ref{Figure:Mfinaldiag} and Table~\ref{table:WD_pred}). The younger the cluster, the higher M$_\mathrm{ini}$ and M$_\mathrm{WD}$, the differences are  substantial only for a GC age variation that is larger than 1~Gyr.

\subsection{He-rich stars predicted by the original FRMS scenario}\label{inihelium}

In Paper III, we reconstructed the $Y_\mathrm{ini}$ distribution, thanks to the theoretical Na-He correlation predicted by the FRMS scenario, to reproduce the observed sodium distribution in NGC~6752 (Fig.~4 in Paper III). Here we use the same initial helium distribution and focus on the WD properties outcome in the same GC. We assume an identical IMF for 1P and 2P stars and we follow the evolution of a population of 300 000 stars with initial masses between 0.3 and 1~M$_\odot$. \\ 

\begin{figure*}[ht]
   \centering
   \includegraphics[width=0.45\textwidth]{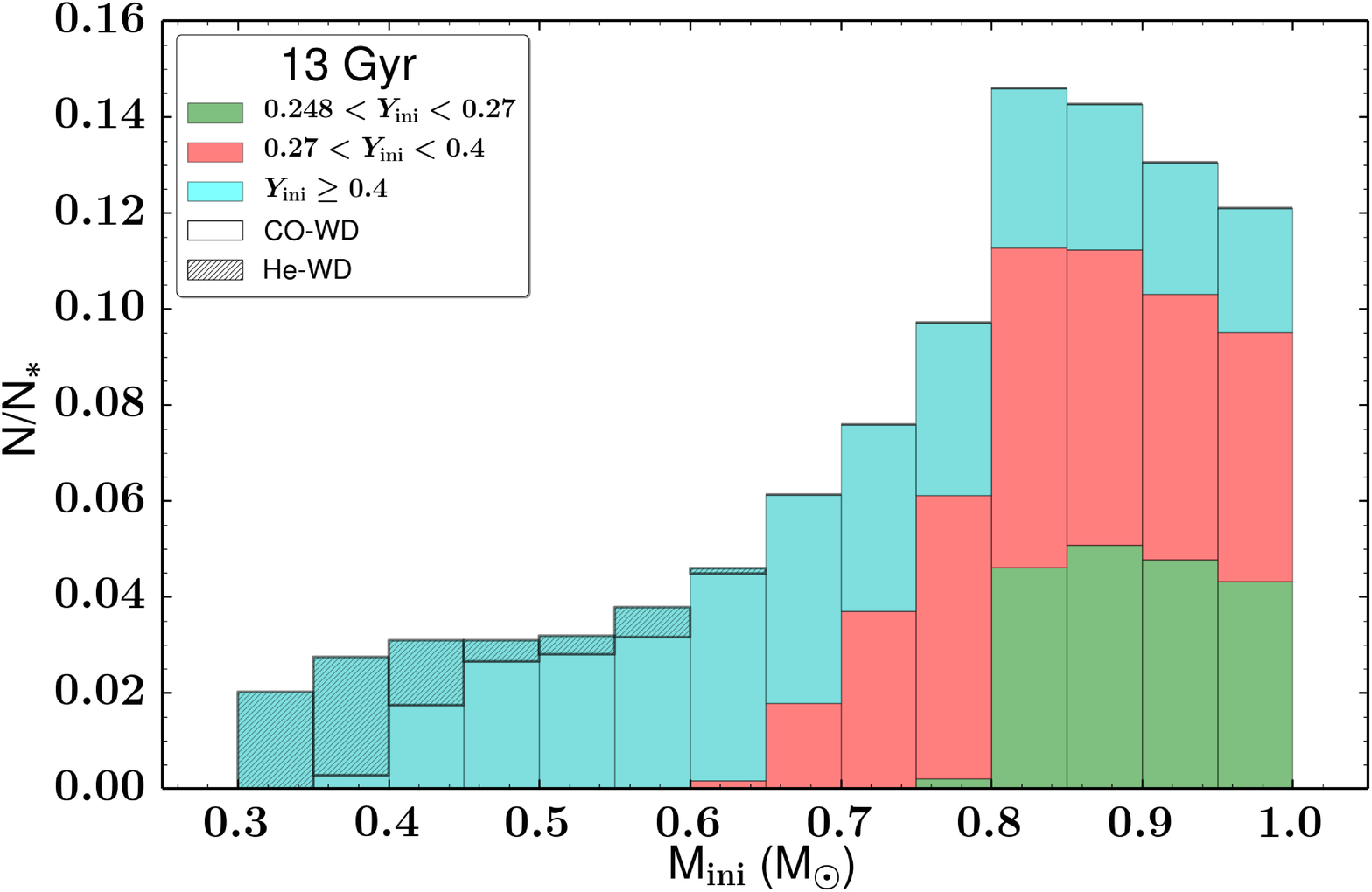}  
   \includegraphics[width=0.45\textwidth]{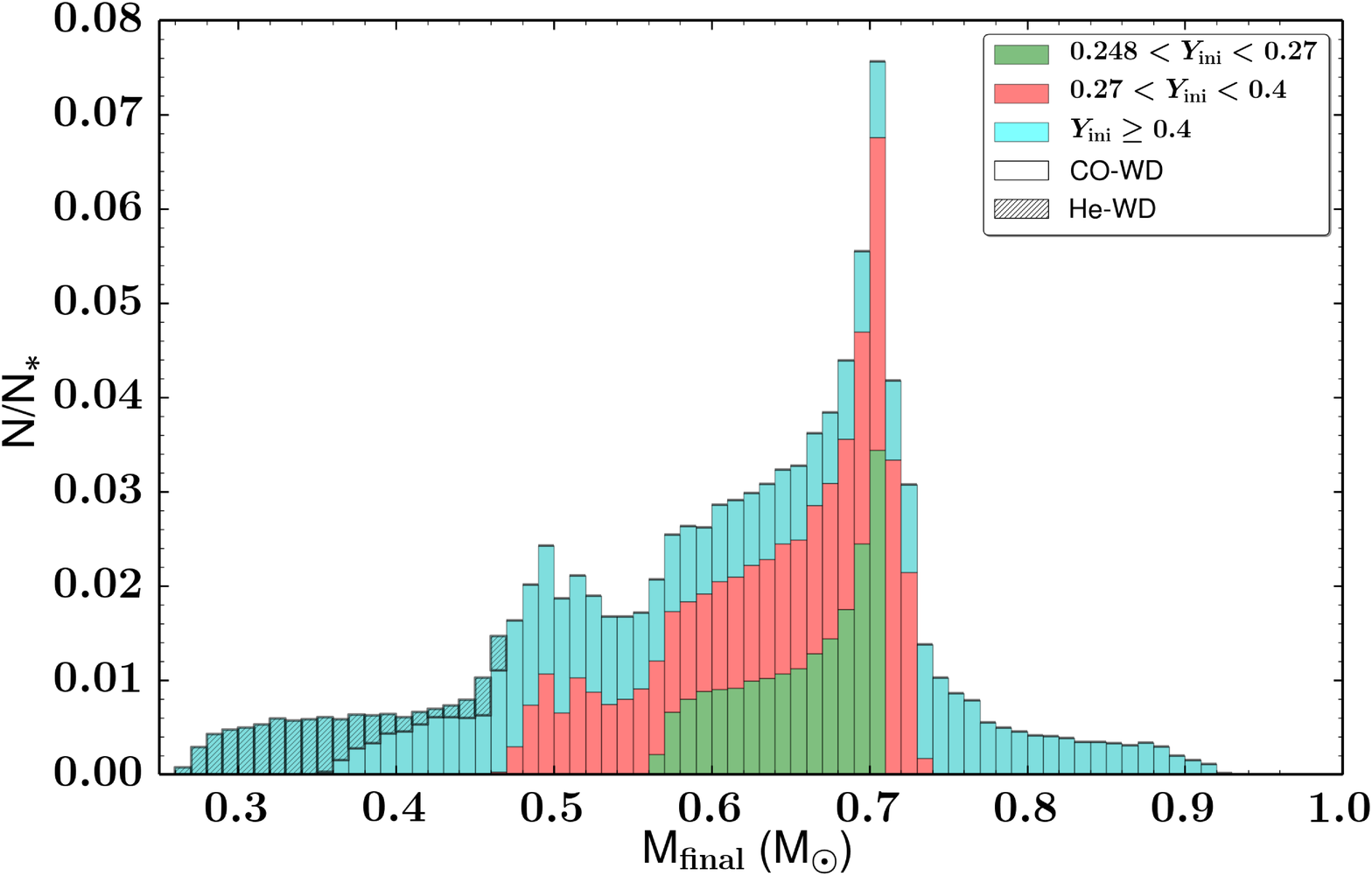} \\
   \includegraphics[width=0.45\textwidth]{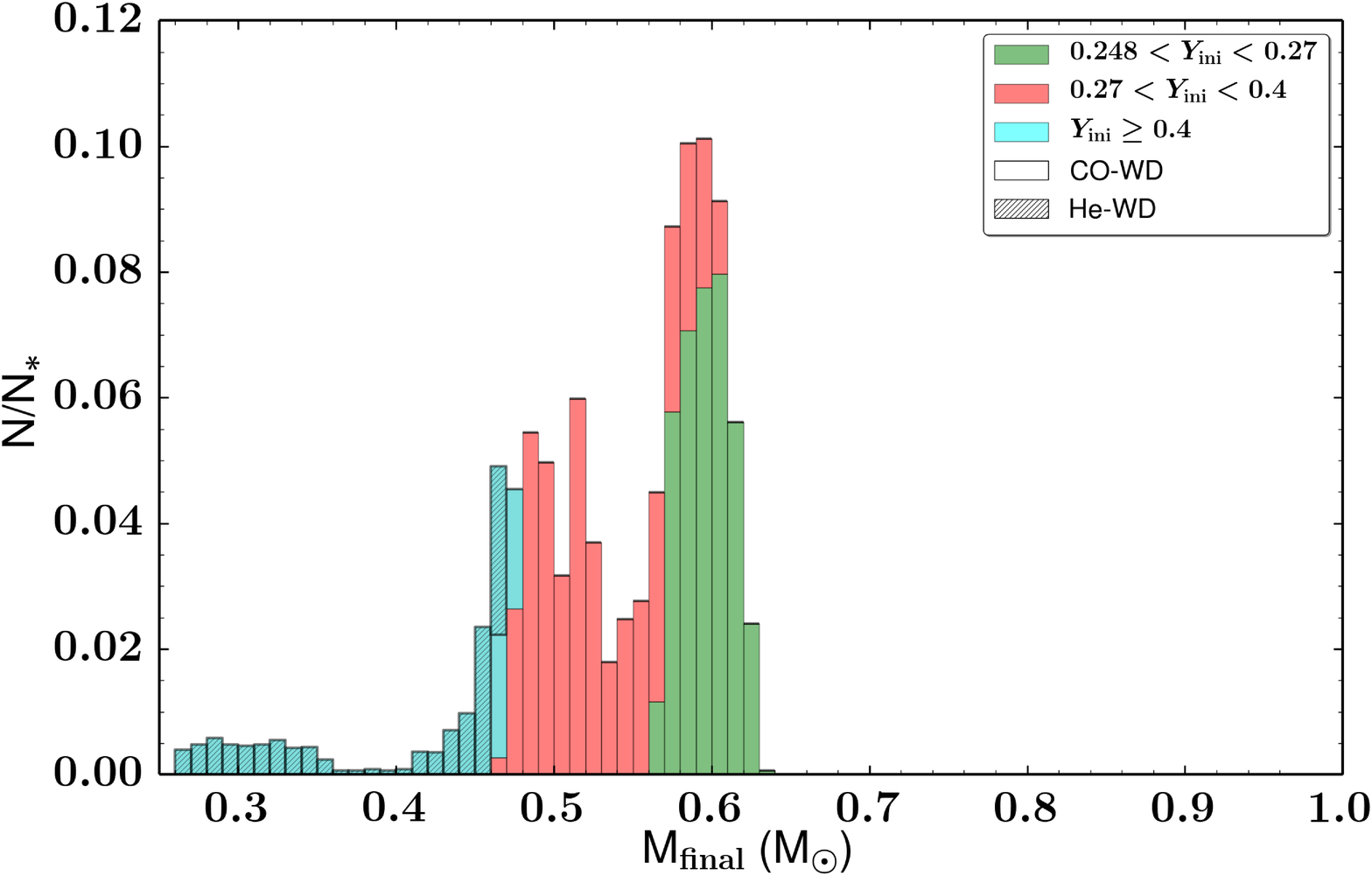} 
   \includegraphics[width=0.45\textwidth]{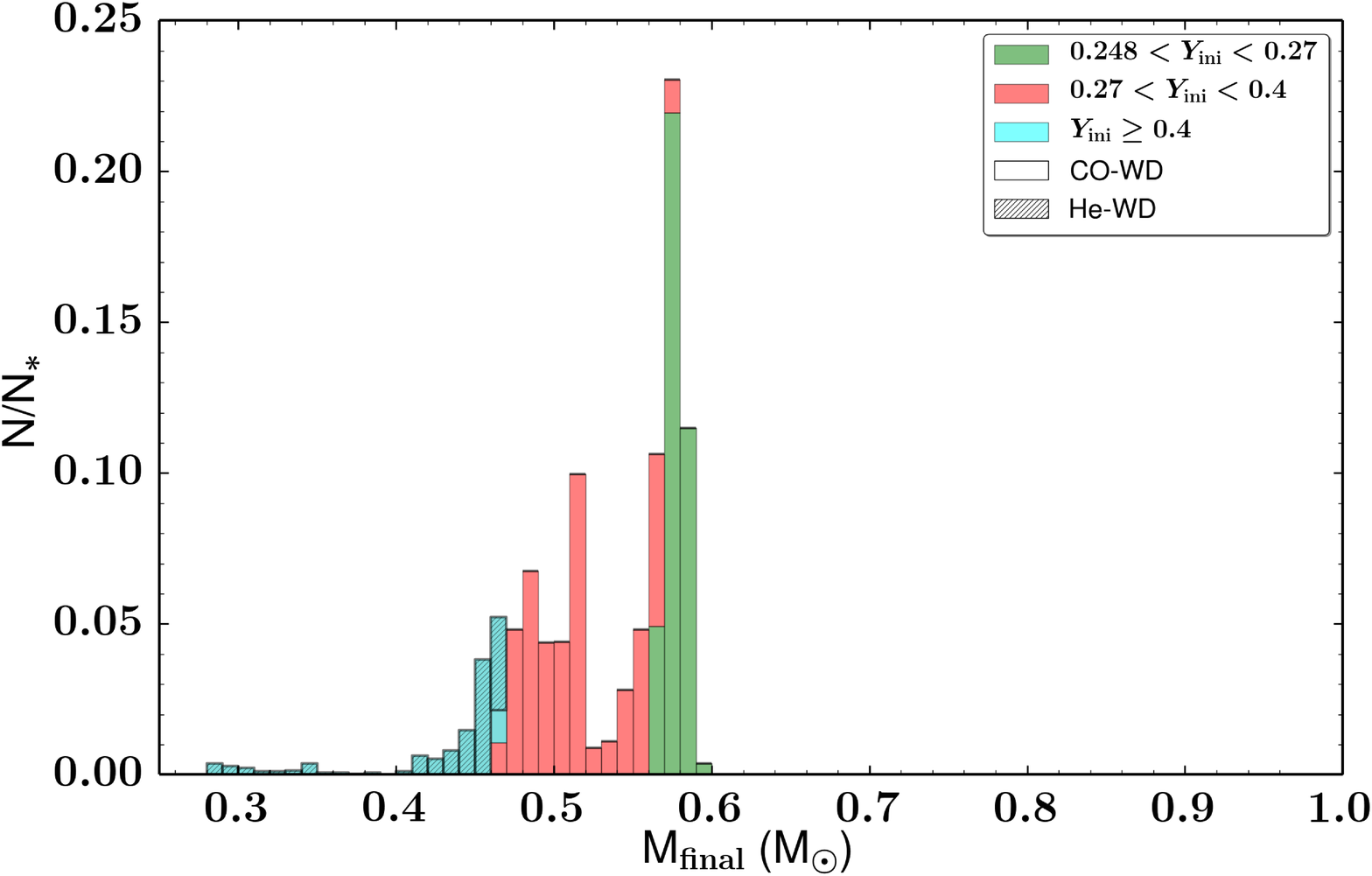}
   \caption{\textit{Top left panel}: Histogram of all the WDs formed as a function of M$_\mathrm{ini}$ (between 0.3 and 1~M$_\odot$) and $Y_\mathrm{ini}$ at 13~Gyr in a model that reproduces the Na distribution in NGC~6752 within the FRMS framework discussed in Paper~III. \textit{Top right panel}: Histogram of the final masses of all the WDs  at 13~Gyr. \textit{Bottom left panel}: WDs present on the WD cooling curve for less than 2~Gyr (upper part of the WD cooling curve). \textit{Bottom right panel}: WDs cooling for less than 500~Myr after the end of the core He burning (brighter part of the WD cooling curve).}
    \label{Figure:WD}
\end{figure*} 

\begin{table*}
\centering
\begin{tabular}{|c |c | c | c | c | c | c | c | c | c | c |}
        \hline
    & \multicolumn{4}{|c|}{13~Gyr} & \multicolumn{4}{|c|}{12~Gyr} \\ \hline
    $Y_\mathrm{ini}$ & N ($\%$) & M ($\%$) & M$_\mathrm{WD}$ domain & M$_\mathrm{WD,mean}$ & N ($\%$) & M ($\%$) & M$_\mathrm{WD}$ domain & M$_\mathrm{WD,mean}$ \\ \hline
    \multicolumn{9}{|c|}{All WDs} \\ \hline
    0.248 & - & - & 0.58-0.70 & 0.66 & - & - & 0.61-0.70 & 0.67 \\
    {[}0.248;0.27] & 19.0 & 20.5 & 0.56-0.71 & 0.66 & 17.9 & 19.4 & 0.58-0.71 & 0.67 \\
    {[}0.27;0.4] & 35.1 & 36.5 & 0.45-0.73 & 0.63 & 34.4 & 36.0 & 0.47-0.73 & 0.64 \\
    $\geq0.4$ & 38.5 & 38.8 & 0.35-0.92 & 0.61 & 40.7 & 40.6 & 0.35-0.92 & 0.61 \\ 
    He-WD & 7.4 & 4.3 & 0.26-0.47 & 0.35 & 7.0 & 4.0 & 0.26-0.47 & 0.35 \\ \hline
    \multicolumn{9}{|c|}{2~Gyr - WDs on the WD cooling curve's upper part} \\ \hline
    0.248 & - & - & 0.58-0.63 & 0.61 & - & - & 0.61-0.66 & 0.63 \\
    {[}0.248;0.27] & 37.8 & 42.0 & 0.57-0.63 & 0.60 & 38.0 & 42.6 & 0.59-0.65 & 0.62 \\
    {[}0.27;0.4] & 46.0 & 45.4 & 0.47-0.61 & 0.53 & 45.0 & 44.6 & 0.47-0.63 & 0.55 \\
    $\geq0.4$ & 3.9 & 3.4 & 0.46-0.48 & 0.47 & 7.1 & 6.1 & 0.46-0.49 & 0.47 \\ 
    He-WD & 12.3 & 9.1 & 0.26-0.47 & 0.40 & 10.0 & 6.7 & 0.26-0.47 & 0.38 \\ \hline
    \multicolumn{9}{|c|}{500~Myr- WDs on the WD cooling curve's brighter part} \\ \hline
    0.248 & - & - & 0.58-0.59 & 0.59 & - & - & 0.61-0.62 & 0.61 \\
    {[}0.248;0.27] & 38.7 & 42.3 & 0.57-0.59 & 0.58 & 39.8 & 44.0 & 0.59-0.62 & 0.60 \\
    {[}0.27;0.4] & 47.9 & 46.7 & 0.47-0.58 & 0.52 & 44.9 & 44.1 & 0.47-0.60 & 0.54 \\
    $\geq0.4$ & 1.1 & 1.0 & 0.46-0.47 & 0.47 & 5.1 & 4.4 & 0.47-0.47 & 0.47 \\ 
    He-WD & 12.3 & 10.0 & 0.28-0.46 & 0.43 & 10.1 & 7.5 & 0.27-0.47 & 0.41 \\     
\hline 
\end{tabular} 
\caption{Proportion of WDs from 0.3-1~M$_\odot$ progenitors (initially 300 000 stars), their mass domain, and their mean value as a function of the $Y_\mathrm{ini}$ domain at 12 and 13~Gyr. The WDs are divided into three subgroups: WDs that joined the cooling sequence  less than 500~Myr ago after the end of the core He burning (brighter part of the WD cooling curve), WDs that joined the cooling sequence for less than 2~Gyr (upper part of the WD cooling curve), and a group with all the WDs.} \label{table:WD_stat}
\end{table*} 

We show in Fig.~\ref{Figure:WD} the number of predicted WDs at 13~Gyr with initial masses ranging between 0.3 and 1~M$_\odot$. For each initial mass bin, different colors are used depending on the initial helium content of the WD progenitors. The value $Y_\mathrm{ini}$ = 0.27 corresponds to $\Delta Y \sim$0.02; this is the helium enrichment found in most GCs \citep[e.g., in NGC~6752;][]{Nardiello15}. The value $Y_\mathrm{ini}$ = 0.4 represents the maximum He enrichment determined in GCs such as NGC~2808 \citep{Milone15} and $\omega$ Centauri \citep{King12}. The shaded areas correspond to He-WDs.
One of the most noticeable results is that He-rich stars with initial mass lower than 0.82~M$_\odot$ (minimum initial mass to form WDs for $Y_\mathrm{ini}$ = 0.248) contribute to increase the number of WDs. At 13~Gyr, 34~\% of the stars are present under the form of WDs under the FRMS assumptions compared to 16~\% if only He-normal stars are considered. The limit CO-WD/He-WD is (M$_\mathrm{ini}$;$Y_\mathrm{ini}$) = (0.62;0.40) (Fig.~2 from paper III). This limit means that at 13~Gyr only stars with M$_\mathrm{ini}$ lower than 0.62~M$_\odot$ and $Y_\mathrm{ini}$ higher than 0.40 are present under the form of He-WD. 

In the top right and bottom panels of Fig.~\ref{Figure:WD}, we show the predicted distribution of WD masses in NGC~6752 and their composition; we list the corresponding values in Table~\ref{table:WD_stat}. 
He-rich stars contribute to an increase of the total mass locked in the GC at 13~Gyr of 91~\% compared to the case with only He-normal stars. At this age, most of the WDs come from progenitors with a non-negligible He enrichment ($Y_\mathrm{ini}\geq$ 0.270), and the minimum mass (initial and current) of CO-WDs significantly decreases for the most He-enriched stars. Finally He-WDs are expected to represent only a very small fraction of WDs (7~\% in number, 4~\% in mass with a mean mass around 0.35~M$_\odot$). Here, these He-WDs are predicted only by the He-rich stars\footnote{No close binary evolution has been considered here; it may produce He-WDs from He-normal stars.}.\\
If we focus on the WDs present on the WD cooling curve for less than 2~Gyr, the CO-WD mean mass is lower compared to the mean mass of all the WDs from progenitors with M$_\mathrm{ini}$ between 0.3 and 1~M$_\odot$, and He-WDs represent ~10~\% of the remnants. \\
The masses and relative proportions of CO- and He-WDs do not change significantly for the WDs that reached the cooling sequence for less than 2~Gyr and less than 500~Myr ago. 
The stars with an initial helium mass fraction higher than 0.4 do not contribute significantly to the proportion of WDs and these stars mainly produce He-WDs.\\
The results at 12~Gyr are not significantly different in term of proportions; the mean M$_\mathrm{WD}$ at a given $Y_\mathrm{ini}$ is slightly higher for the three categories of WDs considered. \\ 
\section{Comparison with observations}\label{compobs}

\begin{table*}
\centering
\begin{tabular}{|c |c | c | c | c | c | c |}
        \hline
    GC ID & \multicolumn{2}{c|}{Observations} & \multicolumn{2}{c|}{Models} & He-WD\\
    & M$_\mathrm{WD}$ & $\Delta Y$ & M$_\mathrm{WD}$ with $Y_\mathrm{canonical}$ & M$_\mathrm{WD}$ with $\Delta Y$ & \\ \hline
        NGC~6752 & 0.53 or 0.59 & 0.03 & 0.59 & [0.55-0.59] & No \\    
        {[}Fe/H] = -1.75 & \cite{Moehler04} & \cite{Milone13} & & 0.57 & \\ \hline
        NGC~6121 & [0.50-0.59] - 0.53 & 0.02 & 0.57 & [0.55-0.57] & No \\
        {[}Fe/H] = -1.15 & \cite{Kalirai09} & \cite{Nardiello15} & & & \\ \hline
        NGC~2808 & - & 0.15 & - & [0.46-0.59] & Yes \\    
        {[}Fe/H] = -1.15 & & \cite{Milone15} & & & \\ \hline 
    $\omega$ Centauri & - & 0.14 & - & [0.47-0.62] & Yes at \\  
    {[}Fe/H] = -1.75 &  & \cite{King12} &  & & [Fe/H] = -1.15 \\  
\hline 
\end{tabular} 
\caption{Mean WD mass, mass domain (between brackets), and He enhancements in mass fraction determined by observations in GCs. Predictions from our models with $Y_\mathrm{ini}$ = $Y_\mathrm{canonical}$ and with He enhancements corresponding to observations for WD mass (mass domain and mean values). The last column lists the possibility of obtaining He-WD from single He-rich progenitors.} \label{table:obs}
\end{table*} 

In this section we compare our results and the determined WD masses in GCs (see Table~\ref{table:obs}). For such a comparison to be valid, we need to compare the characteristics of WDs deduced from our model at an age equivalent to that of the cluster. To be consistent we should determine the age of the cluster using our models, but as shown above the results do not vary significantly between 12 and 13~Gyr, therefore, we make our comparisons with an age of 13~Gyr for NGC~6752 \citep[upper limit for this GC; see, e.g.,][]{Gratton03,McDonald15}. Finally, in a  comparison with observations, we use WDs present on the cooling curve for less than 500~Myr since they correspond to the brighter WDs, which are those usually observed in GCs.

\subsection{NGC~6752}

In NGC~6752, \cite{Moehler04} found an average mass for 12  among the brightest isolated WDs of 0.53~M$_\odot$, assuming a distance modulus $(m-M)_0$ = 13.20, or 0.59~M$_\odot$ , if $(m-M)_0$ = 13.05. This mass is very sensitive to this distance modulus; with the updated value of $(m-M)_0 =$ 13.13 from \cite{Harris96} (2010 edition) for this GC, one would expect that the average mass for WDs would be between 0.53~M$_\odot$ and 0.59~M$_\odot$. At 13~Gyr, our models predict a mean value of 0.59~M$_\odot$ when considering the canonical initial helium content (c.f. Table~\ref{table:WD_stat}), which is compatible with the upper value of the observed range. 
This value changes only by -0.01~M$_\odot$ if we also take into account stars slightly He enriched ($0.248 \leq Y_\mathrm{ini} \leq 0.270$). 

\cite{Milone13} determined He variations among NGC~6752 main sequence stars of 0.03 in mass fraction (maximum $Y_\mathrm{ini}$ of $\sim$0.280). At 13~Gyr, our models predict a  mean mass of WDs of 0.57~M$_\odot$, which is very close to the He-normal case. In addition we only expect to have CO-WD from single progenitors because of the low He enrichment. At 12~Gyr, these values are only slightly higher with a mean WD mass of 0.60~M$_\odot$ (for stars with $0.248 \leq Y_\mathrm{ini} \leq 0.280$). 

Instead of the initial He distribution predicted by the FRMS, as a simple test we assume an initial helium distribution with three peaks at $Y_\mathrm{ini}$ = 0.248, 0.258, and 0.278 (25, 45, and 30~\%, respectively, of the total number of stars). These peaks correspond to the three populations identified by \cite{Milone13} on the main sequence in NGC~6752 (populations $a$, $b,$ and $c,$ respectively). In this case, 82~\% of stars are still alive at 13~Gyr. The WDs mean masses of the brighter part of the cooling curve is 0.57~M$_\odot$, thus there are only negligible differences with the He-normal case.

Finally assuming an He enrichment up to $Y_\mathrm{ini}$ = 0.4 enlarges the domain of masses for WDs down to 0.47~M$_\odot$. It would then be interesting to increase the observational samples to better constrain the WD mass domain. This would be an indirect but complementary constraint on the He enrichment.\\

\subsection{NGC~6121}

\cite{Kalirai09} derived the spectroscopic mass of six~WDs located near the tip of the cooling sequence in NGC~6121 \citep[{[}Fe/H{]} = -1.16,][2010 edition]{Harris96}. They found a mean mass of 0.53$\pm$0.01~M$_\odot$ with a range between 0.50~M$_\odot$ and 0.59~M$_\odot$. For this GC we adopt an age at the turn-off of 12~Gyr \citep{Hansen04}, an alpha-enhancement of +0.3 and [Fe/H] = -1.15. We predict that He-normal stars at the end of the central He-burning phase have an initial mass of 0.88~M$_\odot$ and produce WD masses of 0.57~M$_\odot$; this theoretical value agrees well with the observational constraints. 

In this GC, \cite{Nardiello15} determined a helium spread of $\Delta Y \sim 0.02$  among main sequence stars. This spread corresponds to $Y_\mathrm{ini} \sim 0.270$, which leads in turn to WDs with a mass of 0.55~M$_\odot$, that is to say, a WD mass variation of 0.02~M$_\odot$, which is lower than the observed range. It means that the dispersion in He content might be larger than that indirectly deduced from observations or that there are physical ingredients in the stellar models that should be improved. To make more insightful comparisons, a larger observational sample is required. Additionally, one should provide the theoretical WD mass distribution as a function of luminosity along the cooling sequence. \\

\subsection{NGC~2808} 
 
NGC~2808 \citep{VandenBerg13}, which is 11~Gyr in age, is  at the same metallicity as NGC~6121, and displays one of the highest He enrichment among GCs. \cite{Bragaglia10} and \cite{Milone15} derived a maximum helium content of $\sim$0.4. In this case the WD masses are between 0.46~M$_\odot$ and 0.59~M$_\odot$. Since the CO-WD/He-WD limit is M$_\mathrm{ini}$ = 0.74~M$_\odot$ and $Y_\mathrm{ini} = 0.35$ (cf. paper II), our models predict that He-WDs from single He-rich progenitors are compatible with the high He enrichment present in NGC~2808. It would then be very interesting to investigate WD masses and compositions in this cluster. \\

\subsection{$\omega$ Centauri} 

$\omega$ Centauri displays a [Fe/H] distribution between -1.83 and -0.42, with a large number of stars around [Fe/H] $\sim$ -1.7 and -1.34 \citep{Villanova14}. This GC is very interesting since observations from \cite{Calamida08} support the possible presence of an unusually large number of He-WDs. \cite{Bellini13} also found in $\omega$ Centauri a double WD cooling sequence and argue that one of them would contain up to $\sim$90~\% of He-core WDs.

We assume the initial helium distribution we took for NGC~6752; at 12~Gyr \citep[$\omega$ Centauri turn-off age;][]{MarinFranch09} and at [Fe/H] = -1.75. $\omega$ Centauri displays a high He enrichment up to $Y_\mathrm{ini}\sim$0.4 \citep[$\Delta Y \sim 0.14$;][]{King12}, and in this case our models predict CO-WD masses down to 0.47~M$_\odot$. The CO-WD/He-WD limit is M$_\mathrm{ini}$ = 0.61~M$_\odot$ and $Y_\mathrm{ini} = 0.43$, thus we do not expect He-WD due to single He-rich progenitors. At [Fe/H] = -1.15\footnote{In this simple test we do not take into account the age spread of the different subpopulations of $\sim$2-4~Gyr found in $\omega$ Centauri \citep{Stanford06}.} , which is the closest metallicity available in our grid from [Fe/H] = -1.34, the limit is M$_\mathrm{ini}$ = 0.75~M$_\odot$ and $Y_\mathrm{ini} = 0.33$. Thus our models also  predict the presence of He-WDs from single He-rich progenitors for this GC \citep[see similar results in][]{Cassisi09}. It is then mandatory to develop the observational determination of WD masses and a full theoretical study of WDs in this GC (taking metallicity, age, and helium spreads into account at the same time).

The high helium enrichment leads to a large decrease of WD masses and in turn to a significant extent of the WD mass domain at a given age. Thus it would be interesting to increase the number of WD masses observationally determined to better constrain the WD mass domain and then provide an additional way to constrain He content in GCs. Finally only NGC~2808 and $\omega$ Centauri are supposed to host He-WDs from single He-rich progenitors among the GCs investigated here.

\section{Conclusion}\label{Conclusions}

We investigated for the first time the impact of the initial distribution of helium content predicted by the FRMS scenario for 1P and 2P stars on the mass of the WD remnants for low-mass, low-metallicity stars. 

We assumed the same mass-loss prescription for He-normal and He-rich models because the mass-loss dependence on the initial helium content is currently unknown. This is one of the weaknesses of the current stellar models since WD masses are very sensitive to the mass-loss prescription. Thus it seems mandatory to further investigate the mass-loss dependence on He content and quantify it at large scales (i.e., among several GCs).

We have shown that high He enrichment leads at an advanced age to very low WD masses and eventually to the formation of He-WDs from single progenitors within GCs lifetime. When the initial helium distribution predicted by the original FRMS scenario is assumed, there is a high increase of the total number and mass of WDs, which might affect the  dynamical properties of the host GCs \citep[see, e.g.,][]{Kruijssen08}.

The modest He enrichment currently estimated by indirect methods in several GCs, if representative of the initial helium distribution, implies that the WD populations are weakly affected by differences in the helium content. However there are GCs with high He enrichment such as NGC~2808 and $\omega$ Centauri, where the WD masses are greatly affected. We also showed that NGC~2808 and $\omega$ Centauri could host He-WD remnants from single He-rich progenitors. Nonetheless,  $\omega$ Centauri might be a remnant core of a disrupted dwarf galaxy and in this case it might be hazardous to compare it to other GCs \citep[e.g.,][]{Ideta04,Bekki06}, thus a full study of this system is mandatory. 

It would be interesting to investigate WD masses in other GCs with high He enrichment, such as  NGC~2419 \citep[$\Delta Y \sim 0.11$;][]{diCriscienzo15}, NGC~6388 \citep[$\Delta Y \sim 0.14$;][]{Busso07} and to a certain extent NGC~6266 \citep[$\Delta Y \sim 0.08$;][]{Milone15_6266}, NGC~6441 \citep[$\Delta Y \sim 0.06$;][]{Bellini13} and NGC~7089 \citep[$\Delta Y \sim 0.07$;][]{Milone15_7089}. The WD mass domain determined would then provide an additional indirect way to measure and constrain the amount of He enrichment in various GCs. 

\begin{acknowledgements}
We acknowledge support from the Swiss National Science Foundation (FNS) for the Project 200020-159543 ``Multiple stellar populations in massive star clusters – Formation, evolution, dynamics, impact on galactic evolution" (PI C.C.). We thank the International Space Science Institute (ISSI, Bern, CH) for welcoming the activities of ISSI Team 271 ``Massive star clusters across the Hubble Time'' (2013 - 2016, team leader C.C.). Finally, the referee Achim Weiss is warmly thanked for his constructive approach, pertinent questions, and suggestions that have greatly helped us improve the presentation of our results. 
\end{acknowledgements}

\bibliographystyle{aa}
\bibliography{Chantereau17}

\begin{thebibliography}{87}
\expandafter\ifx\csname natexlab\endcsname\relax\def\natexlab#1{#1}\fi

\bibitem[{{Althaus} {et~al.}(2017){Althaus}, {De Ger{\'o}nimo}, {C{\'o}rsico},
  {Torres}, \& {Garc{\'{\i}}a-Berro}}]{Althaus16}
{Althaus}, L.~G., {De Ger{\'o}nimo}, F., {C{\'o}rsico}, A., {Torres}, S., \&
  {Garc{\'{\i}}a-Berro}, E. 2017, \aap, 597, A67

\bibitem[{{Anderson} {et~al.}(2009){Anderson}, {Piotto}, {King}, {Bedin}, \&
  {Guhathakurta}}]{Anderson09}
{Anderson}, J., {Piotto}, G., {King}, I.~R., {Bedin}, L.~R., \& {Guhathakurta},
  P. 2009, \apjl, 697, L58

\bibitem[{{Bastian} {et~al.}(2015){Bastian}, {Cabrera-Ziri}, \&
  {Salaris}}]{Bastian15}
{Bastian}, N., {Cabrera-Ziri}, I., \& {Salaris}, M. 2015, \mnras, 449, 3333

\bibitem[{{Bedin} {et~al.}(2005){Bedin}, {Salaris}, {Piotto}, {King},
  {Anderson}, {Cassisi}, \& {Momany}}]{Bedin05}
{Bedin}, L.~R., {Salaris}, M., {Piotto}, G., {et~al.} 2005, \apjl, 624, L45

\bibitem[{{Bekki} \& {Norris}(2006)}]{Bekki06}
{Bekki}, K. \& {Norris}, J.~E. 2006, \apjl, 637, L109

\bibitem[{{Bellini} {et~al.}(2013){Bellini}, {Anderson}, {Salaris}, {Cassisi},
  {Bedin}, {Piotto}, \& {Bergeron}}]{Bellini13}
{Bellini}, A., {Anderson}, J., {Salaris}, M., {et~al.} 2013, \apjl, 769, L32

\bibitem[{{Bertelli} {et~al.}(2008){Bertelli}, {Girardi}, {Marigo}, \&
  {Nasi}}]{Bertelli08}
{Bertelli}, G., {Girardi}, L., {Marigo}, P., \& {Nasi}, E. 2008, \aap, 484, 815

\bibitem[{{Bowen} \& {Willson}(1991)}]{Bowen91}
{Bowen}, G.~H. \& {Willson}, L.~A. 1991, \apjl, 375, L53

\bibitem[{{Bragaglia} {et~al.}(2010{\natexlab{a}}){Bragaglia}, {Carretta},
  {Gratton}, {D'Orazi}, {Cassisi}, \& {Lucatello}}]{Bragaglia10}
{Bragaglia}, A., {Carretta}, E., {Gratton}, R., {et~al.} 2010{\natexlab{a}},
  \aap, 519, A60

\bibitem[{{Bragaglia} {et~al.}(2010{\natexlab{b}}){Bragaglia}, {Carretta},
  {Gratton}, {Lucatello}, {Milone}, {Piotto}, {D'Orazi}, {Cassisi}, {Sneden},
  \& {Bedin}}]{Bragaglia10_2808}
{Bragaglia}, A., {Carretta}, E., {Gratton}, R.~G., {et~al.} 2010{\natexlab{b}},
  \apjl, 720, L41

\bibitem[{{Brown} {et~al.}(2001){Brown}, {Sweigart}, {Lanz}, {Landsman}, \&
  {Hubeny}}]{Brown01}
{Brown}, T.~M., {Sweigart}, A.~V., {Lanz}, T., {Landsman}, W.~B., \& {Hubeny},
  I. 2001, \apj, 562, 368

\bibitem[{{Busso} {et~al.}(2007){Busso}, {Cassisi}, {Piotto}, {Castellani},
  {Romaniello}, {Catelan}, {Djorgovski}, {Recio Blanco}, {Renzini}, {Rich},
  {Sweigart}, \& {Zoccali}}]{Busso07}
{Busso}, G., {Cassisi}, S., {Piotto}, G., {et~al.} 2007, \aap, 474, 105

\bibitem[{{Calamida} {et~al.}(2008){Calamida}, {Corsi}, {Bono}, {Stetson},
  {Prada Moroni}, {Degl'Innocenti}, {Ferraro}, {Iannicola}, {Koester},
  {Pulone}, {Monelli}, {Amico}, {Buonanno}, {Caputo}, {D'Odorico},
  {Freyhammer}, {Marchetti}, {Nonino}, \& {Romaniello}}]{Calamida08}
{Calamida}, A., {Corsi}, C.~E., {Bono}, G., {et~al.} 2008, \apjl, 673, L29

\bibitem[{{Carretta} {et~al.}(2009){Carretta}, {Bragaglia}, {Gratton}, \&
  {Lucatello}}]{Carretta09}
{Carretta}, E., {Bragaglia}, A., {Gratton}, R., \& {Lucatello}, S. 2009, \aap,
  505, 139

\bibitem[{{Carretta} {et~al.}(2010){Carretta}, {Bragaglia}, {Gratton},
  {Recio-Blanco}, {Lucatello}, {D'Orazi}, \& {Cassisi}}]{Carretta10}
{Carretta}, E., {Bragaglia}, A., {Gratton}, R.~G., {et~al.} 2010, \aap, 516,
  A55

\bibitem[{{Cassisi} {et~al.}(2009){Cassisi}, {Salaris}, {Anderson}, {Piotto},
  {Pietrinferni}, {Milone}, {Bellini}, \& {Bedin}}]{Cassisi09}
{Cassisi}, S., {Salaris}, M., {Anderson}, J., {et~al.} 2009, \apj, 702, 1530

\bibitem[{{Castellani} \& {Castellani}(1993)}]{Castellani93}
{Castellani}, M. \& {Castellani}, V. 1993, \apj, 407, 649

\bibitem[{{Chantereau} {et~al.}(2015){Chantereau}, {Charbonnel}, \&
  {Decressin}}]{Chantereau15}
{Chantereau}, W., {Charbonnel}, C., \& {Decressin}, T. 2015, \aap, 578, A117

\bibitem[{{Chantereau} {et~al.}(2016){Chantereau}, {Charbonnel}, \&
  {Meynet}}]{Chantereau16}
{Chantereau}, W., {Charbonnel}, C., \& {Meynet}, G. 2016, \aap, 592, A111

\bibitem[{{Charbonnel}(2016)}]{Charbonnel16_EES}
{Charbonnel}, C. 2016, in EAS Publications Series, Vol.~80, EAS Publications
  Series, ed. E.~{Moraux}, Y.~{Lebreton}, \& C.~{Charbonnel}, 177--226

\bibitem[{{Charbonnel} \& {Chantereau}(2016)}]{Charbonnel16}
{Charbonnel}, C. \& {Chantereau}, W. 2016, \aap, 586, A21

\bibitem[{{D'Cruz} {et~al.}(1996){D'Cruz}, {Dorman}, {Rood}, \&
  {O'Connell}}]{DCruz96}
{D'Cruz}, N.~L., {Dorman}, B., {Rood}, R.~T., \& {O'Connell}, R.~W. 1996, \apj,
  466, 359

\bibitem[{{Decressin} {et~al.}(2007){Decressin}, {Meynet}, {Charbonnel},
  {Prantzos}, \& {Ekstr{\"o}m}}]{Decressin07}
{Decressin}, T., {Meynet}, G., {Charbonnel}, C., {Prantzos}, N., \&
  {Ekstr{\"o}m}, S. 2007, \aap, 464, 1029

\bibitem[{{di Criscienzo} {et~al.}(2010{\natexlab{a}}){di Criscienzo},
  {D'Antona}, \& {Ventura}}]{diCriscienzo10_6397}
{di Criscienzo}, M., {D'Antona}, F., \& {Ventura}, P. 2010{\natexlab{a}}, \aap,
  511, A70

\bibitem[{{Di Criscienzo} {et~al.}(2015){Di Criscienzo}, {Tailo}, {Milone},
  {D'Antona}, {Ventura}, {Dotter}, \& {Brocato}}]{diCriscienzo15}
{Di Criscienzo}, M., {Tailo}, M., {Milone}, A.~P., {et~al.} 2015, \mnras, 446,
  1469

\bibitem[{{di Criscienzo} {et~al.}(2010{\natexlab{b}}){di Criscienzo},
  {Ventura}, {D'Antona}, {Milone}, \& {Piotto}}]{diCriscienzo10}
{di Criscienzo}, M., {Ventura}, P., {D'Antona}, F., {Milone}, A., \& {Piotto},
  G. 2010{\natexlab{b}}, \mnras, 408, 999

\bibitem[{{Doherty} {et~al.}(2014){Doherty}, {Gil-Pons}, {Lau}, {Lattanzio}, \&
  {Siess}}]{Doherty14}
{Doherty}, C.~L., {Gil-Pons}, P., {Lau}, H.~H.~B., {Lattanzio}, J.~C., \&
  {Siess}, L. 2014, \mnras, 437, 195

\bibitem[{{Dupree} \& {Avrett}(2013)}]{Dupree13}
{Dupree}, A.~K. \& {Avrett}, E.~H. 2013, \apjl, 773, L28

\bibitem[{{Forestini} \& {Charbonnel}(1997)}]{Forestini97}
{Forestini}, M. \& {Charbonnel}, C. 1997, \aaps, 123

\bibitem[{{Garc{\'{\i}}a-Berro} {et~al.}(2014){Garc{\'{\i}}a-Berro}, {Torres},
  {Althaus}, \& {Miller Bertolami}}]{GarciaBerro14}
{Garc{\'{\i}}a-Berro}, E., {Torres}, S., {Althaus}, L.~G., \& {Miller
  Bertolami}, M.~M. 2014, \aap, 571, A56

\bibitem[{{Gratton} {et~al.}(2003){Gratton}, {Bragaglia}, {Carretta},
  {Clementini}, {Desidera}, {Grundahl}, \& {Lucatello}}]{Gratton03}
{Gratton}, R.~G., {Bragaglia}, A., {Carretta}, E., {et~al.} 2003, \aap, 408,
  529

\bibitem[{{Gratton} {et~al.}(2012){Gratton}, {Carretta}, \&
  {Bragaglia}}]{Gratton12}
{Gratton}, R.~G., {Carretta}, E., \& {Bragaglia}, A. 2012, \aapr, 20, 50

\bibitem[{{Gratton} {et~al.}(2014){Gratton}, {Lucatello}, {Sollima},
  {Carretta}, {Bragaglia}, {Momany}, {D'Orazi}, {Cassisi}, \&
  {Salaris}}]{Gratton14}
{Gratton}, R.~G., {Lucatello}, S., {Sollima}, A., {et~al.} 2014, \aap, 563, A13

\bibitem[{{Gratton} {et~al.}(2015){Gratton}, {Lucatello}, {Sollima},
  {Carretta}, {Bragaglia}, {Momany}, {D'Orazi}, {Salaris}, {Cassisi}, \&
  {Stetson}}]{Gratton15}
{Gratton}, R.~G., {Lucatello}, S., {Sollima}, A., {et~al.} 2015, \aap, 573, A92

\bibitem[{{Groenewegen} {et~al.}(2009){Groenewegen}, {Sloan}, {Soszy{\'n}ski},
  \& {Petersen}}]{Groenewegen09}
{Groenewegen}, M.~A.~T., {Sloan}, G.~C., {Soszy{\'n}ski}, I., \& {Petersen},
  E.~A. 2009, \aap, 506, 1277

\bibitem[{{Groenewegen} {et~al.}(1998){Groenewegen}, {Whitelock}, {Smith}, \&
  {Kerschbaum}}]{Groenewegen98}
{Groenewegen}, M.~A.~T., {Whitelock}, P.~A., {Smith}, C.~H., \& {Kerschbaum},
  F. 1998, \mnras, 293, 18

\bibitem[{{Hansen}(2005)}]{Hansen05}
{Hansen}, B.~M.~S. 2005, \apj, 635, 522

\bibitem[{{Hansen} {et~al.}(2003){Hansen}, {Kalogera}, \& {Rasio}}]{Hansen03}
{Hansen}, B.~M.~S., {Kalogera}, V., \& {Rasio}, F.~A. 2003, \apj, 586, 1364

\bibitem[{{Hansen} \& {Liebert}(2003)}]{Hansen03_review}
{Hansen}, B.~M.~S. \& {Liebert}, J. 2003, \araa, 41, 465

\bibitem[{{Hansen} {et~al.}(2004){Hansen}, {Richer}, {Fahlman}, {Stetson},
  {Brewer}, {Currie}, {Gibson}, {Ibata}, {Rich}, \& {Shara}}]{Hansen04}
{Hansen}, B.~M.~S., {Richer}, H.~B., {Fahlman}, G.~G., {et~al.} 2004, \apjs,
  155, 551

\bibitem[{{Harris}(1996)}]{Harris96}
{Harris}, W.~E. 1996, \aj, 112, 1487

\bibitem[{{Ideta} \& {Makino}(2004)}]{Ideta04}
{Ideta}, M. \& {Makino}, J. 2004, \apjl, 616, L107

\bibitem[{{Kalirai} {et~al.}(2009){Kalirai}, {Saul Davis}, {Richer},
  {Bergeron}, {Catelan}, {Hansen}, \& {Rich}}]{Kalirai09}
{Kalirai}, J.~S., {Saul Davis}, D., {Richer}, H.~B., {et~al.} 2009, \apj, 705,
  408

\bibitem[{{King} {et~al.}(2012){King}, {Bedin}, {Cassisi}, {Milone}, {Bellini},
  {Piotto}, {Anderson}, {Pietrinferni}, \& {Cordier}}]{King12}
{King}, I.~R., {Bedin}, L.~R., {Cassisi}, S., {et~al.} 2012, \aj, 144, 5

\bibitem[{{Kruijssen} \& {Lamers}(2008)}]{Kruijssen08}
{Kruijssen}, J.~M.~D. \& {Lamers}, H.~J.~G.~L.~M. 2008, \aap, 490, 151

\bibitem[{{Lagarde} {et~al.}(2012){Lagarde}, {Decressin}, {Charbonnel},
  {Eggenberger}, {Ekstr{\"o}m}, \& {Palacios}}]{Lagarde12}
{Lagarde}, N., {Decressin}, T., {Charbonnel}, C., {et~al.} 2012, \aap, 543,
  A108

\bibitem[{{Lee} {et~al.}(2013){Lee}, {Han}, {Joo}, {Jang}, {Na}, {Okamoto},
  {Arimoto}, {Lim}, {Kim}, \& {Yoon}}]{Lee13}
{Lee}, Y.-W., {Han}, S.-I., {Joo}, S.-J., {et~al.} 2013, \apjl, 778, L13

\bibitem[{{Mar{\'{\i}}n-Franch} {et~al.}(2009){Mar{\'{\i}}n-Franch},
  {Aparicio}, {Piotto}, {Rosenberg}, {Chaboyer}, {Sarajedini}, {Siegel},
  {Anderson}, {Bedin}, {Dotter}, {Hempel}, {King}, {Majewski}, {Milone},
  {Paust}, \& {Reid}}]{MarinFranch09}
{Mar{\'{\i}}n-Franch}, A., {Aparicio}, A., {Piotto}, G., {et~al.} 2009, \apj,
  694, 1498

\bibitem[{{Marino} {et~al.}(2014){Marino}, {Milone}, {Przybilla}, {Bergemann},
  {Lind}, {Asplund}, {Cassisi}, {Catelan}, {Casagrande}, {Valcarce}, {Bedin},
  {Cort{\'e}s}, {D'Antona}, {Jerjen}, {Piotto}, {Schlesinger}, {Zoccali}, \&
  {Angeloni}}]{Marino14}
{Marino}, A.~F., {Milone}, A.~P., {Przybilla}, N., {et~al.} 2014, \mnras, 437,
  1609

\bibitem[{{McDonald} \& {Zijlstra}(2015)}]{McDonald15}
{McDonald}, I. \& {Zijlstra}, A.~A. 2015, \mnras, 448, 502

\bibitem[{{Miller Bertolami} {et~al.}(2008){Miller Bertolami}, {Althaus},
  {Unglaub}, \& {Weiss}}]{MillerBertolami08}
{Miller Bertolami}, M.~M., {Althaus}, L.~G., {Unglaub}, K., \& {Weiss}, A.
  2008, \aap, 491, 253

\bibitem[{{Milone}(2015)}]{Milone15_6266}
{Milone}, A.~P. 2015, \mnras, 446, 1672

\bibitem[{{Milone} {et~al.}(2013){Milone}, {Marino}, {Piotto}, {Bedin},
  {Anderson}, {Aparicio}, {Bellini}, {Cassisi}, {D'Antona}, {Grundahl},
  {Monelli}, \& {Yong}}]{Milone13}
{Milone}, A.~P., {Marino}, A.~F., {Piotto}, G., {et~al.} 2013, \apj, 767, 120

\bibitem[{{Milone} {et~al.}(2015{\natexlab{a}}){Milone}, {Marino}, {Piotto},
  {Bedin}, {Anderson}, {Renzini}, {King}, {Bellini}, {Brown}, {Cassisi},
  {D'Antona}, {Jerjen}, {Nardiello}, {Salaris}, {Marel}, {Vesperini}, {Yong},
  {Aparicio}, {Sarajedini}, \& {Zoccali}}]{Milone15_7089}
{Milone}, A.~P., {Marino}, A.~F., {Piotto}, G., {et~al.} 2015{\natexlab{a}},
  \mnras, 447, 927

\bibitem[{{Milone} {et~al.}(2015{\natexlab{b}}){Milone}, {Marino}, {Piotto},
  {Renzini}, {Bedin}, {Anderson}, {Cassisi}, {D'Antona}, {Bellini}, {Jerjen},
  {Pietrinferni}, \& {Ventura}}]{Milone15}
{Milone}, A.~P., {Marino}, A.~F., {Piotto}, G., {et~al.} 2015{\natexlab{b}},
  \apj, 808, 51

\bibitem[{{Moehler} \& {Bono}(2008)}]{Moehler08}
{Moehler}, S. \& {Bono}, G. 2008, ArXiv e-prints

\bibitem[{{Moehler} {et~al.}(2007){Moehler}, {Dreizler}, {Lanz}, {Bono},
  {Sweigart}, {Calamida}, {Monelli}, \& {Nonino}}]{Moehler07}
{Moehler}, S., {Dreizler}, S., {Lanz}, T., {et~al.} 2007, \aap, 475, L5

\bibitem[{{Moehler} {et~al.}(2004){Moehler}, {Koester}, {Zoccali}, {Ferraro},
  {Heber}, {Napiwotzki}, \& {Renzini}}]{Moehler04}
{Moehler}, S., {Koester}, D., {Zoccali}, M., {et~al.} 2004, \aap, 420, 515

\bibitem[{{Mucciarelli} {et~al.}(2014){Mucciarelli}, {Lovisi}, {Lanzoni}, \&
  {Ferraro}}]{Mucciarelli14}
{Mucciarelli}, A., {Lovisi}, L., {Lanzoni}, B., \& {Ferraro}, F.~R. 2014, \apj,
  786, 14

\bibitem[{{Nardiello} {et~al.}(2015){Nardiello}, {Milone}, {Piotto}, {Marino},
  {Bellini}, \& {Cassisi}}]{Nardiello15}
{Nardiello}, D., {Milone}, A.~P., {Piotto}, G., {et~al.} 2015, \aap, 573, A70

\bibitem[{{Paresce} \& {De Marchi}(2000)}]{Paresce00}
{Paresce}, F. \& {De Marchi}, G. 2000, \apj, 534, 870

\bibitem[{{Pasquini} {et~al.}(2011){Pasquini}, {Mauas}, {K{\"a}ufl}, \&
  {Cacciari}}]{Pasquini11}
{Pasquini}, L., {Mauas}, P., {K{\"a}ufl}, H.~U., \& {Cacciari}, C. 2011, \aap,
  531, A35

\bibitem[{{Pietrinferni} {et~al.}(2009){Pietrinferni}, {Cassisi}, {Salaris},
  {Percival}, \& {Ferguson}}]{Pietrinferni09}
{Pietrinferni}, A., {Cassisi}, S., {Salaris}, M., {Percival}, S., \&
  {Ferguson}, J.~W. 2009, \apj, 697, 275

\bibitem[{{Piotto}(2009)}]{Piotto09}
{Piotto}, G. 2009, in IAU Symposium, Vol. 258, The Ages of Stars, ed. E.~E.
  {Mamajek}, D.~R. {Soderblom}, \& R.~F.~G. {Wyse}, 233--244

\bibitem[{{Piotto} {et~al.}(2007){Piotto}, {Bedin}, {Anderson}, {King},
  {Cassisi}, {Milone}, {Villanova}, {Pietrinferni}, \& {Renzini}}]{Piotto07}
{Piotto}, G., {Bedin}, L.~R., {Anderson}, J., {et~al.} 2007, \apjl, 661, L53

\bibitem[{{Reimers}(1975)}]{Reimers75}
{Reimers}, D. 1975, Memoires of the Societe Royale des Sciences de Liege, 8,
  369

\bibitem[{{Renzini} {et~al.}(1996){Renzini}, {Bragaglia}, {Ferraro},
  {Gilmozzi}, {Ortolani}, {Holberg}, {Liebert}, {Wesemael}, \&
  {Bohlin}}]{Renzini96}
{Renzini}, A., {Bragaglia}, A., {Ferraro}, F.~R., {et~al.} 1996, \apjl, 465,
  L23

\bibitem[{{Renzini} {et~al.}(2015){Renzini}, {D'Antona}, {Cassisi}, {King},
  {Milone}, {Ventura}, {Anderson}, {Bedin}, {Bellini}, {Brown}, {Piotto}, {van
  der Marel}, {Barbuy}, {Dalessandro}, {Hidalgo}, {Marino}, {Ortolani},
  {Salaris}, \& {Sarajedini}}]{Renzini15}
{Renzini}, A., {D'Antona}, F., {Cassisi}, S., {et~al.} 2015, \mnras, 454, 4197

\bibitem[{{Richer} {et~al.}(2013){Richer}, {Goldsbury}, {Heyl}, {Hurley},
  {Dotter}, {Kalirai}, {Woodley}, {Fahlman}, {Rich}, \& {Shara}}]{Richer13}
{Richer}, H.~B., {Goldsbury}, R., {Heyl}, J., {et~al.} 2013, \apj, 778, 104

\bibitem[{{Salaris} {et~al.}(2006){Salaris}, {Weiss}, {Ferguson}, \&
  {Fusilier}}]{Salaris06}
{Salaris}, M., {Weiss}, A., {Ferguson}, J.~W., \& {Fusilier}, D.~J. 2006, \apj,
  645, 1131

\bibitem[{{Salpeter}(1955)}]{Salpeter55}
{Salpeter}, E.~E. 1955, \apj, 121, 161

\bibitem[{{Sbordone} {et~al.}(2011){Sbordone}, {Salaris}, {Weiss}, \&
  {Cassisi}}]{Sbordone11}
{Sbordone}, L., {Salaris}, M., {Weiss}, A., \& {Cassisi}, S. 2011, \aap, 534,
  A9

\bibitem[{{Schr{\"o}der} \& {Cuntz}(2005)}]{Schroder05}
{Schr{\"o}der}, K.-P. \& {Cuntz}, M. 2005, \apjl, 630, L73

\bibitem[{{Siess} {et~al.}(2000){Siess}, {Dufour}, \& {Forestini}}]{Siess00}
{Siess}, L., {Dufour}, E., \& {Forestini}, M. 2000, \aap, 358, 593

\bibitem[{{Stanford} {et~al.}(2006){Stanford}, {Da Costa}, {Norris}, \&
  {Cannon}}]{Stanford06}
{Stanford}, L.~M., {Da Costa}, G.~S., {Norris}, J.~E., \& {Cannon}, R.~D. 2006,
  \apj, 647, 1075

\bibitem[{{Strickler} {et~al.}(2009){Strickler}, {Cool}, {Anderson}, {Cohn},
  {Lugger}, \& {Serenelli}}]{Strickler09}
{Strickler}, R.~R., {Cool}, A.~M., {Anderson}, J., {et~al.} 2009, \apj, 699, 40

\bibitem[{{Torres} {et~al.}(2015){Torres}, {Garc{\'{\i}}a-Berro}, {Althaus}, \&
  {Camisassa}}]{Torres15}
{Torres}, S., {Garc{\'{\i}}a-Berro}, E., {Althaus}, L.~G., \& {Camisassa},
  M.~E. 2015, \aap, 581, A90

\bibitem[{{Valcarce} {et~al.}(2012){Valcarce}, {Catelan}, \&
  {Sweigart}}]{Valcarce12}
{Valcarce}, A.~A.~R., {Catelan}, M., \& {Sweigart}, A.~V. 2012, \aap, 547, A5

\bibitem[{{VandenBerg} {et~al.}(2013){VandenBerg}, {Brogaard}, {Leaman}, \&
  {Casagrande}}]{VandenBerg13}
{VandenBerg}, D.~A., {Brogaard}, K., {Leaman}, R., \& {Casagrande}, L. 2013,
  \apj, 775, 134

\bibitem[{{Vassiliadis} \& {Wood}(1993)}]{Vassiliadis93}
{Vassiliadis}, E. \& {Wood}, P.~R. 1993, \apj, 413, 641

\bibitem[{{Ventura} {et~al.}(2001){Ventura}, {D'Antona}, {Mazzitelli}, \&
  {Gratton}}]{Ventura01}
{Ventura}, P., {D'Antona}, F., {Mazzitelli}, I., \& {Gratton}, R. 2001, \apjl,
  550, L65

\bibitem[{{Ventura} {et~al.}(2013){Ventura}, {Di Criscienzo}, {Carini}, \&
  {D'Antona}}]{Ventura13}
{Ventura}, P., {Di Criscienzo}, M., {Carini}, R., \& {D'Antona}, F. 2013,
  \mnras, 431, 3642

\bibitem[{{Villanova} {et~al.}(2014){Villanova}, {Geisler}, {Gratton}, \&
  {Cassisi}}]{Villanova14}
{Villanova}, S., {Geisler}, D., {Gratton}, R.~G., \& {Cassisi}, S. 2014, \apj,
  791, 107

\bibitem[{{Villanova} {et~al.}(2012){Villanova}, {Geisler}, {Piotto}, \&
  {Gratton}}]{Villanova12}
{Villanova}, S., {Geisler}, D., {Piotto}, G., \& {Gratton}, R.~G. 2012, \apj,
  748, 62

\bibitem[{{Villanova} {et~al.}(2009){Villanova}, {Piotto}, \&
  {Gratton}}]{Villanova09}
{Villanova}, S., {Piotto}, G., \& {Gratton}, R.~G. 2009, \aap, 499, 755

\bibitem[{{Weidemann}(2000)}]{Weidemann00}
{Weidemann}, V. 2000, \aap, 363, 647

\bibitem[{{Zoccali} {et~al.}(2001){Zoccali}, {Renzini}, {Ortolani},
  {Bragaglia}, {Bohlin}, {Carretta}, {Ferraro}, {Gilmozzi}, {Holberg},
  {Marconi}, {Rich}, \& {Wesemael}}]{Zoccali01}
{Zoccali}, M., {Renzini}, A., {Ortolani}, S., {et~al.} 2001, \apj, 553, 733

\end{thebibliography}

\end{document}